# Correlation between magnetic domain structures and quantum anomalous Hall effect in epitaxial MnBi$_2$Te$_4$ thin films

Yang Shi[1]*, Yunhe Bai[2]*, Yuanzhao Li[2,3]*, Yang Feng[3]*, Qiang Li[1], Huanyu Zhang[1], Yang Chen[2], Yitian Tong[2], Jianli Luan[2,3], Ruixuan Liu[2], Pengfei Ji[2], Zongwei Gao[3], Hangwen Guo[1,6], Jinsong Zhang[2,4,5], Yayu Wang[2,4,5], Xiao Feng[2,3,4,5]†, Ke He[2,3,4,5], Xiaodong Zhou[1,6]†, and Jian Shen[1,6,7]†

[1]State Key Laboratory of Surface Physics, Institute for Nanoelectronic Devices and Quantum Computing, and Department of Physics, Fudan University, Shanghai 200433, China.
[2]State Key Laboratory of Low Dimensional Quantum Physics, Department of Physics, Tsinghua University, Beijing 100084, China.
[3]Beijing Academy of Quantum Information Science, Beijing 100193, China
[4]Hefei National Laboratory, Hefei 230088, P. R. China
[5]Frontier Science Center for Quantum Information, Beijing 100084, China.
[6]Zhangjiang Fudan International Innovation Center, Fudan University, Shanghai 201210, China
[7]Shanghai Research Center for Quantum Sciences, Shanghai 201315, China

*These authors contributed equally to this work
†Emails: xiaofeng@mail.tsinghua.edu.cn, zhouxd@fudan.edu.cn, shenj5494@fudan.edu.cn

**Abstract:**

We use magnetic force microscopy (MFM) to study spatial uniformity of magnetization of epitaxially grown MnBi$_2$Te$_4$ thin films. Compared to films which exhibit no quantum anomalous Hall effect (QAH), films with QAH are observed to have more spatial uniformity of magnetization with larger domain size. The domain evolution upon magnetic field sweeping indicates that the magnetic domains or the spatial nonuniformity of magnetization originates from the strong pinning of the inherent sample inhomogeneity. A direct correlation between the Hall resistivity $\rho_{yx}$ and the domain size has been established by analyzing a series of thin films with and without QAH. Our observation shows that one has to suppress the spatial nonuniformity of magnetization to allow the Hall resistivity $\rho_{yx}$ to be quantized. The fact that a sizable longitudinal resistivity $\rho_{xx}$ remains even for the QAH sample suggests a quantized Hall insulator scenario. Our work provides important insights to the understanding of the quantization mechanism and the dissipation of the QAH state in MnBi$_2$Te$_4$ system.

A robust quantum anomalous Hall (QAH) state, i.e., the quantized Hall effect in the absence of external magnetic field, is highly desirable for its potential applications in energy-efficient electronic devices and topological quantum computing. After the first experimental observation of QAH in a magnetically doped topological insulator (TI) [1], QAH has been recently discovered in both the intrinsic magnetic TI MnBi$_2$Te$_4$ [2] and Moire superlattices of two-dimensional materials [3-7]. Compared to magnetically doped TI with inherent dopant-induced disorder, the stoichiometric MnBi$_2$Te$_4$ should have significantly less disorder and is thus expected to be a cleaner platform to realize QAH at a higher temperature. While most of work so far have been devoted to studying QAH of MnBi$_2$Te$_4$ thin flakes exfoliated from single crystals [2,8-12], efforts have been made to grow epitaxial MnBi$_2$Te$_4$ thin films using molecular beam epitaxy (MBE), given its advantage for device applications with high reproducibility and scalability [13-22].

Y. Bai *et al.* reported a quantized Hall resistivity in MBE-grown MnBi$_2$Te$_4$ thin films under high magnetic fields but with an insulating ground state at zero field [23]. By optimizing the sample growth, they have recently achieved the Hall quantization at zero field, i.e., the QAH state [24]. However, a large longitudinal resistivity ($\rho_{xx} \sim 1\ h/e^2$) still exists at zero field, incompatible with a clean QAH state with dissipationless chiral edge state. The MBE samples are thus ideal to investigate the Hall quantization mechanism and the origin of dissipation in the QAH state of MnBi$_2$Te$_4$. Specifically, one would like to know how the system evolves from an insulating ground state to the QAH state at zero field, and what remains to cause a large dissipation in the QAH state.

In this work, we conduct a magnetic force microscopy (MFM) study of MBE-grown MnBi$_2$Te$_4$ thin films with and without QAH to provide a magnetic perspective into the issue. MFM imaging shows that both QAH and non-QAH samples are magnetically non-uniform, although the QAH sample displays a significantly improved spatial uniformity of magnetization compared to that of the non-QAH samples. Specifically, the QAH sample exhibits larger magnetic domain size and a distinct field-induced metamagnetic transition. The field-dependent measurements indicate that the origin of the magnetic nonuniformity is tied to strong domain pinning effect caused by the sample's inherent inhomogeneity. As a consequence, instead of a homogenous QAH phase, the system should be considered as a random network of QAH puddles surrounded by dissipationless chiral edge states. The longitudinal transport happens via the electron tunneling between neighboring puddles, leading to a sizable dissipation.

MnBi$_2$Te$_4$ is an intrinsic antiferromagnetic (AFM) TI consisting of Te-Bi-Te-Mn-Te-Bi-Te septuple layers (SLs) that are antiferromagnetically coupled along the out-of-plane direction, forming an A-type AFM order [25-27]. Odd-SL MnBi$_2$Te$_4$ with an uncompensated magnetization is predicted to be a QAH state. An external magnetic field can drive an AFM to ferromagnetic (FM) phase transition resulting in a FM state at high fields. Prior to the MFM measurements, transport measurements were carried out on the MBE-grown samples and five 5-SL MnBi$_2$Te$_4$ samples with different transport quantization levels were selected for MFM study as schematically shown in Fig. 1(a). Figure 1(b) shows their field-dependent Hall resistivity $\rho_{yx}(\mu_0 H)$ at above 1.6 K. Note that samples #1 to #4 are from the same batch as those reported in the earlier work [23], which only displays the Hall quantization at high fields but not at zero field. On the other hand, sample #5 is from the latest work and realizes the Hall quantization at zero field [24]. In Fig. 1(b), all samples exhibit a clear magnetic hysteresis loop with the coercive field $H_c \sim 1$ T suggesting a FM behavior due to the uncompensated magnetization in odd-layer samples. Noticeably, sample #5 shows a significant boost in the zero field Hall resistivity compared to the four other samples. We refer sample #5 (samples #1 to #4) to the zero-field QAH sample (non-QAH samples) hereafter.

MFM measurements indicate that all the non-QAH samples (#1 to #4) show similar characteristics whereas the QAH sample (#5) behaves differently. Figure 2 shows a series of selected MFM images taken on sample #3 to illustrate the typical magnetic properties of the non-QAH samples (see supplementary 1 for the full data set). The sample was first zero-field cooled to 6 K, and MFM images were then taken as snapshots upon field ramping. The top row of Fig. 2 corresponds to the initial magnetization process. At zero field, the sample is in a multi-domain state (Fig. 2a). The domain pattern evolves with the increasing field and the roughness of the MFM image is reduced up to 2 T (Fig. 2d), consistent with the trend that the sample is polarized towards a single domain even though the MFM signal still shows some spatial variations at 2 T. As the field further increases, the image becomes rough again (Fig. 2e) and forms a different domain pattern at 8.2 T with a stronger spatial variation of MFM signal (Fig. 2g). The central row of Fig. 2 shows the domain evolution as the field is reduced to zero. The complex domain pattern at 8.2 T gradually disappears at 2 T (Fig. 2k). The sample goes back to a relatively uniform state at 0 T (Fig. 2n), indicating that the sample tends to become a single domain state. The field is subsequently applied along the opposite direction (bottom row of Fig. 2). Starting from a relatively uniform state at 0 T, the sample enters into a multi-domain state at -1 T (Fig. 2p) corresponding to the FM domain reversal at the coercive field. The sample returns back to a uniform state at -2 T (Fig. 2r). As the field further increases, the complex domain pattern observed at the positive high fields reappears with identical characteristic features (compare Fig. 2u to Fig. 2g). The domain evolution process is virtually mirror imaged upon field sweeping in the reversed direction from -8.2 T to 8.2 T (supplementary 1).

Figure 3 shows the domain evolution of the QAH sample #5 following identical magnetization process with the same MFM measurement parameters as sample #3 (see supplementary 2 for the full data set), which allows a direct comparison between Figs. 2 and 3. While the field dependence of the domain evolution shows similar characteristics between the QAH and the non-QAH samples, the spatial uniformity of the MFM signal of the QAH sample is significantly improved.

To better compare the domain evolution quantitatively, we calculate the standard deviation of the MFM signal ($\delta f$) for each MFM image and plot it as a function of the magnetic field. $\delta f$ characterizes the spatial uniformity of magnetization. For a typical FM domain evolution, $\delta f$ peaks at the coercive field $\pm H_c$ when up and down domains are equally populated, and diminishes for a single uniform domain state [28]. Figure 4(a) shows the $\delta f(\mu_0 H)$ of sample #3, which is typical for all the non-QAH samples (supplementary 3). At small fields when the 5-SL MnBi$_2$Te$_4$ is in an A-type AFM order, $\delta f$ peaks at $\pm H_c$ (~±1 T) owing to the FM domain reversal. The fact that MFM determined $H_c$ value agrees well with that determined by the Hall measurement (Fig. 1b) suggests that MFM imaging represents the global properties nicely. At high fields when the 5-SL MnBi$_2$Te$_4$ undergoes an AFM to FM phase transition, $\delta f$ continuously increases without saturation. This behavior is likely caused by a spatially nonuniform AFM to FM phase transition over a broad field range which enhances the spatial variation of sample magnetization with increasing field (supplementary 4). As a result, $\delta f$ keeps increasing with the field, in stark contrast to the magnetically doped TI with a diminished $\delta f$ at high fields when the system is magnetized towards a single uniform domain [28].

Figure 4(b) shows the $\delta f(\mu_0 H)$ of QAH sample #5. Similar to sample #3, $\delta f(\mu_0 H)$ peaks at $\pm H_c$ as expected. The absolute value of $\delta f$ in sample #5 is, however, considerably lower than that of sample #3 especially at high fields. This is due to an enhanced spatial uniformity of magnetization in the QAH sample as directly observed in the MFM images (compare Fig. 2 to Fig. 3). Interestingly, $\delta f$ exhibits a second peak at ~±3.5 T in the QAH sample, which likely reflects the spin-flop metamagnetic transition of MnBi$_2$Te$_4$ causing spatial variation of the net magnetization [29]. A similar and even sharper peak of domain contrast at this spin-flop transition was observed in the MFM measurement of MnBi$_2$Te$_4$ single crystal with a spatially uniform MFM signal [30]. Note that such metamagnetic transition induced $\delta f(\mu_0 H)$ peak is not observable in the non-QAH sample because the metamagnetic transition does not occur at the same critical field due to its high spatial nonuniformity. Above the critical field of the metamagnetic transition, $\delta f$ of the QAH sample increases slightly and saturates above 7 T, indicating a nearly saturated though still somewhat spatially nonuniform sample magnetization. In stark contrast, there is no sign of $\delta f$ saturation in the non-QAH sample up to 8 T, implying that the magnetization of the non-QAH sample is not only highly spatially nonuniform but also far away from reaching the saturation.

To shed lights on the transport quantization at zero field (QAH state), we take a closer look at the magnetic state at zero field after the initial magnetization. Although being relatively uniform compared to the multi-domain state at high fields, the MFM image at zero field displays a finite signal spatial variation reflecting the spatial nonuniformity of magnetization. A typical length scale of such variation can be extracted by Fourier transform (FT) analysis of the MFM image [31,32]. Figure 4(c) shows an example of zero field MFM image of sample #5 (the same as Fig. 3n) whose FT image is shown in the inset. Figure 4(d) shows the azimuthally averaged FT spectral intensity as a function of wave vector *k*. We apply a polynomial curve fitting to the envelope of the spectrum and find the peak of the envelope at $\lambda \sim 1.8$ $\mu m$, corresponding to the characteristic length scale of magnetization nonuniformity, or magnetic domain size. Similar analysis was performed for all five samples which unambiguously demonstrates an enhanced spatial uniformity of magnetization with larger domain size in the QAH sample (supplementary 5). Figure 4(e) plots the zero field Hall resistivity $\rho_{yx}(0\ \text{T})$ measured from transport (Fig. 1b) against this magnetic domain size $\lambda(0\ \text{T})$ estimated from MFM. The $\rho_{yx}(0\ \text{T})$ increases nearly linearly with the domain size $\lambda(0\ \text{T})$. Apparently, a boost of the zero field Hall resistivity of the QAH sample is associated with its larger domain size.

The MFM results provide important insights to the understanding of QAH in MnBi$_2$Te$_4$ thin films. In contrast to MnBi$_2$Te$_4$ single crystals showing a large (~10 μm) and uniform magnetic domain [30,33], MnBi$_2$Te$_4$ thin films show small and very inhomogeneous domains. In addition, MFM measurements reveal a very strong domain pinning effect. For example, the domain patterns at the positive and the negative $H_c$ (supplementary 1 and 2), the ones at the positive and the negative high fields (compare Fig. 2g to Fig. 2u), and even the intermediate states at 2 T in up- and down-sweep directions (compare Fig. 2k to Fig. 2r) are all identical to each other. This can only be understood if one assumes that the domain nucleation and growth is closely tied to the heterogeneity of the sample whose origin need further investigations. For example, the magnetization nonuniformity can arise from the spatial variation of magnetic anisotropy which is sensitive to the local strain state of thin films. Chemical disordering, such as the formation of Mn$_{Bi}$ antisite defects, can also cause a spatial variation of effective magnetization [34-36]. This kind of inhomogeneity has strong spatial registry to explain the observed strong domain pinning.

It is also very intriguing to find a linear correlation between $\rho_{yx}(0\ \text{T})$ and $\lambda(0\ \text{T})$. Also note that sample #5 displays a large dissipation ($\rho_{xx} \sim 1\ h/e^2$) in its QAH state [24]. Previous study on two-dimensional electron gas reported a quantum Hall (QH) liquid to Hall insulator transition with the latter displaying a diverging $\rho_{xx}$ but a finite $\rho_{yx}$ as the temperature approaches zero [37-39]. A quantized Hall insulator was later found amid the transition featuring a quantized $\rho_{yx}$ in the insulating state [40]. A semiclassical network model was constructed to explain such a quantized Hall insulator as a random network of QH puddles surrounded by dissipationless chiral edge states in which the charge transport happens via the electron tunneling between the neighboring puddles (Fig. 4f) [41,42]. Most importantly, this model shows that $\rho_{yx}$ can remain quantized if the electron dephasing length is shorter than the typical puddle size. We attribute the QAH state of sample #5 to such a quantized Hall insulator for the following reasons. First, our samples are probably composed of high-quality MnBi$_2$Te$_4$ regions (puddles) exhibiting quantized $\rho_{yx}$ and vanishing $\rho_{xx}$, with boundaries of other deteriorated regions between them. This inhomogeneity likely causes the nonuniformity of magnetization as revealed by MFM. Second, the fact that a boost in $\rho_{yx}(0\ \text{T})$ of sample #5 comes with an increasing $\lambda(0\ \text{T})$ strongly indicates that the Hall quantization is possible as long as the QAH puddle size is large enough exceeding a critical value, i.e., the electron dephasing length in the framework of quantized Hall insulator. Third, the residual $\rho_{xx}$ naturally comes from the electron tunneling between the QAH puddles. Previous work reported a universal $\rho_{xx}$ ($\sim 1\ h/e^2$) at the transition between QH liquid to Hall insulator [43]. The fact that $\rho_{xx}(0\ \text{T})$ of sample #5 is close to such a quantized value further corroborates our speculation of a quantized Hall insulator state.

In summary, MFM study on MnBi$_2$Te$_4$ thin films reveals a spatially inhomogeneous FM state. Such magnetic inhomogeneity comes from the inherent spatial disordering whose exact origin calls for further investigations. A quantized Hall insulator scenario was considered to explain the Hall quantization mechanism as well as the origin of dissipation.

The work at Fudan University is supported by National Key Research Program of China (Grant Nos. 2022YFA1403300, 2020YFA0309100), National Natural Science Foundation of China (Grant Nos. 12074080, 12074073 and 12274088) and Shanghai Municipal Science and Technology Major Project (Grant No 2019SHZDZX01). The work at Tsinghua University is supported by the Innovation Program for Quantum Science and Technology (Grant No. 2021ZD0302502), the Basic Science Center Project of National Natural Science Foundation of China (Grant No. 51788104), National Key Research Program of China (Grant No. 2018YFA0307100), and National Natural Science Foundation of China (Grant Nos. 92065206 and 11904053). This work is supported in part by the Beijing Advanced Innovation Center for Future Chip (ICFC) and the Tencent Foundation.

**Reference:**
[1]   C. Z. Chang, J. S. Zhang, X. Feng, J. Shen, Z. C. Zhang, M. H. Guo, K. Li, Y. B. Ou, P. Wei, L. L. Wang, Z. Q. Ji, Y. Feng, S. H. Ji, X. Chen, J. F. Jia, X. Dai, Z. Fang, S. C. Zhang, K. He, Y. Y. Wang, L. Lu, X. C. Ma, and Q. K. Xue, Experimental Observation of the Quantum Anomalous Hall Effect in a Magnetic Topological Insulator, Science **340**, 167 (2013).
[2]   Y. J. Deng, Y. J. Yu, M. Z. Shi, Z. X. Guo, Z. H. Xu, J. Wang, X. H. Chen, and Y. B. Zhang, Quantum anomalous Hall effect in intrinsic magnetic topological insulator MnBi$_2$Te$_4$, Science **367**, 895 (2020).
[3]   M. Serlin, C. L. Tschirhart, H. Polshyn, Y. Zhang, J. Zhu, K. Watanabe, T. Taniguchi, L. Balents, and A. F. Young, Intrinsic quantized anomalous Hall effect in a moiré heterostructure, Science **367**, 900 (2020).
[4]   G. Chen, A. L. Sharpe, E. J. Fox, Y.-H. Zhang, S. Wang, L. Jiang, B. Lyu, H. Li, K. Watanabe, T. Taniguchi, Z. Shi, T. Senthil, D. Goldhaber-Gordon, Y. Zhang, and F. Wang, Tunable correlated Chern insulator and ferromagnetism in a moiré superlattice, Nature **579**, 56 (2020).
[5]   T. Li, S. Jiang, B. Shen, Y. Zhang, L. Li, Z. Tao, T. Devakul, K. Watanabe, T. Taniguchi, L. Fu, J. Shan, and K. F. Mak, Quantum anomalous Hall effect from intertwined moiré bands, Nature **600**, 641 (2021).

[6]   H. Park, J. Cai, E. Anderson, Y. Zhang, J. Zhu, X. Liu, C. Wang, W. Holtzmann, C. Hu, Z. Liu, T. Taniguchi, K. Watanabe, J.-h. Chu, T. Cao, L. Fu, W. Yao, C.-Z. Chang, D. Cobden, D. Xiao, and X. Xu, Observation of Fractionally Quantized Anomalous Hall Effect, Nature **622**, 74 (2023).

[7]   F. Xu, Z. Sun, T. Jia, C. Liu, C. Xu, C. Li, Y. Gu, K. Watanabe, T. Taniguchi, B. Tong, J. Jia, Z. Shi, S. Jiang, Y. Zhang, X. Liu, and T. Li, Observation of Integer and Fractional Quantum Anomalous Hall Effects in Twisted Bilayer $MoTe_2$, Phys. Rev. X **13**, 031037 (2023).

[8]   C. Liu, Y. C. Wang, H. Li, Y. Wu, Y. X. Li, J. H. Li, K. He, Y. Xu, J. S. Zhang, and Y. Y. Wang, Robust axion insulator and Chern insulator phases in a two-dimensional antiferromagnetic topological insulator, Nat. Mater. **19**, 522 (2020).

[9]   J. Ge, Y. Z. Liu, J. H. Li, H. Li, T. C. Luo, Y. Wu, Y. Xu, and J. Wang, High-Chern-number and high-temperature quantum Hall effect without Landau levels, Natl. Sci. Rev. **7**, 1280 (2020).

[10]  Z. Ying, S. Zhang, B. Chen, B. Jia, F. Fei, M. Zhang, H. Zhang, X. Wang, and F. Song, Experimental evidence for dissipationless transport of the chiral edge state of the high-field Chern insulator in $MnBi_2Te_4$ nanodevices, Phys. Rev. B **105**, 085412 (2022).

[11]  D. Ovchinnikov, X. Huang, Z. Lin, Z. Y. Fei, J. Q. Cai, T. C. Song, M. H. He, Q. N. Jiang, C. Wang, H. Li, Y. Y. Wang, Y. Wu, D. Xiao, J. H. Chu, J. Q. Yan, C. Z. Chang, Y. T. Cui, and X. D. Xu, Intertwined Topological and Magnetic Orders in Atomically Thin Chern Insulator $MnBi_2Te_4$, Nano Lett. **21**, 2544 (2021).

[12]  J. Q. Cai, D. Ovchinnikov, Z. Y. Fei, M. H. He, T. C. Song, Z. Lin, C. Wang, D. Cobden, J. H. Chu, Y. T. Cui, C. Z. Chang, D. Xiao, J. Q. Yan, and X. D. Xu, Electric control of a canted-antiferromagnetic Chern insulator, Nat. Commun. **13**, 1668 (2022).

[13]  Y. Gong, J. W. Guo, J. H. Li, K. J. Zhu, M. H. Liao, X. Z. Liu, Q. H. Zhang, L. Gu, L. Tang, X. Feng, D. Zhang, W. Li, C. L. Song, L. L. Wang, P. Yu, X. Chen, Y. Y. Wang, H. Yao, W. H. Duan, Y. Xu, S. C. Zhang, X. C. Ma, Q. K. Xue, and K. He, Experimental Realization of an Intrinsic Magnetic Topological Insulator, Chin. Phys. Lett. **36**, 076801 (2019).

[14]  P. Chen, Q. Yao, J. Q. Xu, Q. Sun, A. J. Grutter, P. Quarterman, P. P. Balakrishnan, C. J. Kinane, A. J. Caruana, S. Langridge, A. Li, B. Achinuq, E. Heppell, Y. C. Ji, S. S. Liu, B. S. Cui, J. M. Liu, P. Y. Huang, Z. K. Liu, G. Q. Yu, F. X. Xiu, T. Hesjedal, J. Zou, X. D. Han, H. J. Zhang, Y. M. Yang, and X. F. Kou, Tailoring the magnetic exchange interaction in $MnBi_2Te_4$ superlattices via the intercalation of ferromagnetic layers, Nat. Electron. **6**, 18 (2023).

[15]  Y. F. Zhao, L. J. Zhou, F. Wang, G. Wang, T. C. Song, D. Ovchinnikov, H. M. Yi, R. B. Mei, K. Wang, M. H. W. Chan, C. X. Liu, X. D. Xu, and C. Z. Chang, Even-Odd Layer-Dependent Anomalous Hall Effect in Topological Magnet $MnBi_2Te_4$ Thin Films, Nano Lett. **21**, 7691 (2021).

[16]  S. H. Su, J. T. Chang, P. Y. Chuang, M. C. Tsai, Y. W. Peng, M. K. Lee, C. M. Cheng, and J. C. A. Huang, Epitaxial Growth and Structural Characterizations of $MnBi_2Te_4$ Thin Films in Nanoscale, Nanomaterials-Basel **11**, 3322 (2021).

[17]  C. X. Trang, Q. L. Li, Y. F. Yin, J. Hwang, G. Akhgar, I. Di Bernardo, A. Grubisic-Cabo, A. Tadich, M. S. Fuhrer, S. K. Mo, N. V. Medhekar, and M. T. Edmonds, Crossover from 2D Ferromagnetic Insulator to Wide Band Gap Quantum Anomalous Hall Insulator in Ultrathin $MnBi_2Te_4$, Acs Nano **15**, 13444 (2021).

[18]  J. Lapano, L. Nuckols, A. R. Mazza, Y.-Y. Pai, J. Zhang, B. Lawrie, R. G. Moore, G. Eres, H. N. Lee, M.-H. Du, T. Z. Ward, J. S. Lee, W. J. Weber, Y. Zhang, and M. Brahlek, Adsorption-controlled growth of $MnTe(Bi_2Te_3)_n$ by molecular beam epitaxy exhibiting stoichiometry-controlled magnetism, Phys. Rev. Mater. **4**, 111201 (2020).

[19] N. Liu, S. Schreyeck, K. M. Fijalkowski, M. Kamp, K. Brunner, C. Gould, and L. W. Molenkamp, Antiferromagnetic order in MnBi$_2$Te$_4$ films grown on Si(111) by molecular beam epitaxy, J. Cryst. Growth **591**, 126677 (2022).

[20] R. Watanabe, R. Yoshimi, M. Kawamura, Y. Kaneko, K. S. Takahashi, A. Tsukazaki, M. Kawasaki, and Y. Tokura, Enhancement of anomalous Hall effect in epitaxial thin films of intrinsic magnetic topological insulator MnBi$_2$Te$_4$ with Fermi-level tuning, Appl. Phys. Lett. **120**, 031901 (2022).

[21] Shanshan Liu, Jiexiang Yu, Enze Zhang, Zihan Li, Qiang Sun, Yong Zhang, Lun Li, Minhao Zhao, Pengliang Leng, Xiangyu Cao, Jin Zou, Xufeng Kou, Jiadong Zang, and F. Xiu, Gate-tunable Intrinsic Anomalous Hall Effect in Epitaxial MnBi$_2$Te$_4$ Films, arXiv:2110.00540 (2021).

[22] Hyunsue Kim, Mengke Liu, Lisa Frammolino, Yanxing Li, Fan Zhang, Woojoo Lee, Chengye Dong, Yi-Fan Zhao, Guan-Yu Chen, Pin-Jui Hsu, Cui-Zu Chang, Joshua Robinson, Jiaqiang Yan, Xiaoqin Li, Allan H. MacDonald, and C.-K. Shih, Atomistic Control in Molecular Beam Epitaxy Growth of Intrinsic Magnetic Topological Insulator MnBi$_2$Te$_4$, arXiv:2309.05656 (2023).

[23] Y. Bai, Y. Li, J. Luan, R. Liu, W. Song, Y. Chen, P.-F. Ji, Q. Zhang, F. Meng, B. Tong, L. Li, Y. Jiang, Z. Gao, L. Gu, J. Zhang, Y. Wang, Q.-K. Xue, K. He, Y. Feng, and X. Feng, Quantized anomalous Hall resistivity achieved in molecular beam epitaxy-grown MnBi$_2$Te$_4$ thin films, Natl. Sci. Rev., nwad189 (2023).

[24] Yuanzhao Li, Yunhe Bai, Yang Feng, Jianli Luan, Zongwei Gao, Yang Chen, Yitian Tong, Ruixuan Liu, Su Kong Chong, Kang L. Wang, Xiaodong Zhou, Jian Shen, Jinsong Zhang, Yayu Wang, Chui-Zhen Chen, XinCheng Xie, Xiao Feng, Ke He, and Q.-K. Xue, Reentrant quantum anomalous Hall effect in molecular beam epitaxy-grown MnBi$_2$Te$_4$ thin films, arXiv:2401.11450 (2024).

[25] M. M. Otrokov, I. I. Klimovskikh, H. Bentmann, D. Estyunin, A. Zeugner, Z. S. Aliev, S. Gass, A. U. B. Wolter, A. V. Koroleva, A. M. Shikin, M. Blanco-Rey, M. Hoffmann, I. P. Rusinov, A. Y. Vyazovskaya, S. V. Eremeev, Y. M. Koroteev, V. M. Kuznetsov, F. Freyse, J. Sanchez-Barriga, I. R. Amiraslanov, M. B. Babanly, N. T. Mamedov, N. A. Abdullayev, V. N. Zverev, A. Alfonsov, V. Kataev, B. Buchner, E. F. Schwier, S. Kumar, A. Kimura, L. Petaccia, G. Di Santo, R. C. Vidal, S. Schatz, K. Kissner, M. Unzelmann, C. H. Min, S. Moser, T. R. F. Peixoto, F. Reinert, A. Ernst, P. M. Echenique, A. Isaeva, and E. V. Chulkov, Prediction and observation of an antiferromagnetic topological insulator, Nature **576**, 416 (2019).

[26] J. H. Li, Y. Li, S. Q. Du, Z. Wang, B. L. Gu, S. C. Zhang, K. He, W. H. Duan, and Y. Xu, Intrinsic magnetic topological insulators in van der Waals layered MnBi$_2$Te$_4$-family materials, Sci. Adv. **5**, aaw5685 (2019).

[27] D. Q. Zhang, M. J. Shi, T. S. Zhu, D. Y. Xing, H. J. Zhang, and J. Wang, Topological Axion States in the Magnetic Insulator MnBi$_2$Te$_4$ with the Quantized Magnetoelectric Effect, Phys. Rev. Lett. **122**, 206401 (2019).

[28] W. B. Wang, Y. B. Ou, C. Liu, Y. Y. Wang, K. He, Q. K. Xue, and W. D. Wu, Direct evidence of ferromagnetism in a quantum anomalous Hall system, Nat. Phys. **14**, 791 (2018).

[29] S. Q. Yang, X. L. Xu, Y. Z. Zhu, R. R. Niu, C. Q. Xu, Y. X. Peng, X. Cheng, X. H. Jia, Y. Huang, X. F. Xu, J. M. Lu, and Y. Ye, Odd-Even Layer-Number Effect and Layer-Dependent Magnetic Phase Diagrams in MnBi$_2$Te$_4$, Phys. Rev. X **11**, 011003 (2021).

[30] P. M. Sass, J. Kim, D. Vanderbilt, J. Q. Yan, and W. D. Wu, Robust A-Type Order and Spin-Flop Transition on the Surface of the Antiferromagnetic Topological Insulator MnBi$_2$Te$_4$, Phys.

Rev. Lett. **125**, 037201 (2020).

[31] W. B. Wang, J. A. Mundy, C. M. Brooks, J. A. Moyer, M. E. Holtz, D. A. Muller, D. G. Schlom, and W. D. Wu, Visualizing weak ferromagnetic domains in multiferroic hexagonal ferrite thin film, Phys. Rev. B **95**, 134443 (2017).

[32] Tatiana A. Webb, Afrin N. Tamanna, Xiaxin Ding, Jikai Xu, Lia Krusin-Elbaum, Cory R. Dean, Dmitri N. Basov, and A. N. Pasupathy, Tunable magnetic domains in ferrimagnetic $MnSb_2Te_4$, arXiv:2308.16806 (2023).

[33] P. M. Sass, W. B. Ge, J. Q. Yan, D. Obeysekera, J. J. Yang, and W. D. Wu, Magnetic Imaging of Domain Walls in the Antiferromagnetic Topological Insulator $MnBi_2Te_4$, Nano Lett. **20**, 2609 (2020).

[34] M. Garnica, M. M. Otrokov, P. C. Aguilar, I. I. Klimovskikh, D. Estyunin, Z. S. Aliev, I. R. Amiraslanov, N. A. Abdullayev, V. N. Zverev, M. B. Babanly, N. T. Mamedov, A. M. Shikin, A. Arnau, A. L. V. de Parga, E. V. Chulkov, and R. Miranda, Native point defects and their implications for the Dirac point gap at $MnBi_2Te_4$(0001), npj Quantum Mater. **7**, 7 (2022).

[35] Y. Lai, L. Q. Ke, J. Q. Yan, R. D. McDonald, and R. J. McQueeney, Defect-driven ferrimagnetism and hidden magnetization in $MnBi_2Te_4$, Phys. Rev. B **103**, 184429 (2021).

[36] X. L. Xu, S. Q. Yang, H. Wang, R. Guzman, Y. C. Gao, Y. Z. Zhu, Y. X. Peng, Z. H. Zang, M. Xi, S. J. Tian, Y. P. Li, H. C. Lei, Z. C. Luo, J. B. Yang, Y. L. Wang, T. L. Xia, W. Zhou, Y. Huang, and Y. Ye, Ferromagnetic-antiferromagnetic coexisting ground state and exchange bias effects in $MnBi_4Te_7$ and $MnBi_6Te_{10}$, Nat. Commun. **13**, 7646 (2022).

[37] S. Kivelson, D. H. Lee, and S. C. Zhang, Global Phase-Diagram in the Quantum Hall-Effect, Phys. Rev. B **46**, 2223 (1992).

[38] R. L. Willett, H. L. Stormer, D. C. Tsui, L. N. Pfeiffer, K. W. West, and K. W. Baldwin, Termination of the series of fractional quantum hall states at small filling factors, Phys. Rev. B **38**, 7881 (1988).

[39] D. Shahar, D. C. Tsui, M. Shayegan, E. Shimshoni, and S. L. Sondhi, Evidence for charge-flux duality near the quantum hall liquid-to-insulator transition, Science **274**, 589 (1996).

[40] M. Hilke, D. Shahar, S. H. Song, D. C. Tsui, Y. H. Xie, and D. Monroe, Experimental evidence for a two-dimensional quantized Hall insulator, Nature **395**, 675 (1998).

[41] E. Shimshoni and A. Auerbach, Quantized Hall insulator: Transverse and longitudinal transport, Phys. Rev. B **55**, 9817 (1997).

[42] L. P. Pryadko and K. Chaltikian, Network of Edge States: A Random Josephson Junction Array Description, Phys. Rev. Lett. **80**, 584 (1998).

[43] D. Shahar, D. C. Tsui, M. Shayegan, R. N. Bhatt, and J. E. Cunningham, Universal Conductivity at the Quantum Hall Liquid to Insulator Transition, Phys. Rev. Lett. **74**, 4511 (1995).

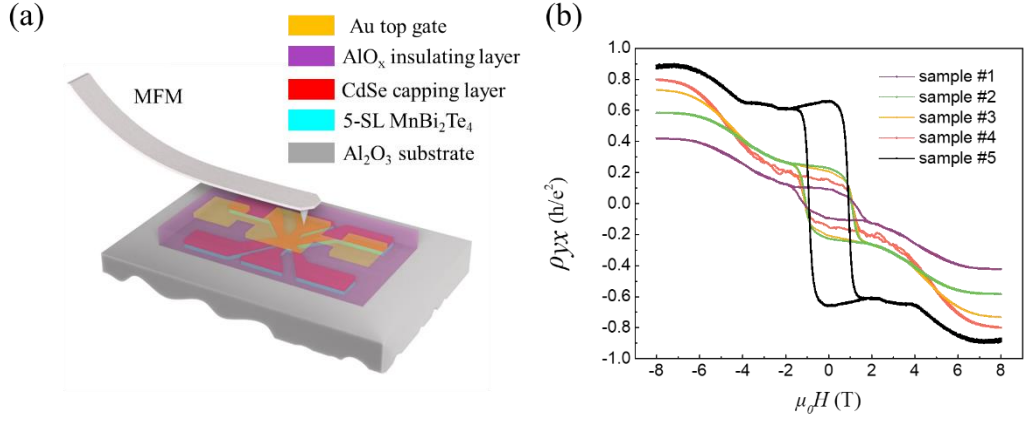

FIG. 1. (a) Schematic of the Hall bar device for MFM and *ex-situ* transport measurement. A 5-SL MnBi$_2$Te$_4$ thin film was grown on the Al$_2$O$_3$ substrate. It was covered by a CdSe capping layer, an AlO$_x$ dielectric layer and a top-gate electrode layer. The top electrode gate and the magnetic tip were grounded during the MFM measurement to eliminate any electrostatic interactions. All MFM images were taken at 6 K in this work. (b) The Hall resistivity $\rho_{yx}(\mu_0 H)$ at best-tuned gate voltage for five samples under this study. Data of samples from #1 to #4 were taken at 1.6 K, whereas the data of sample #5 was taken at 2 K in this work.

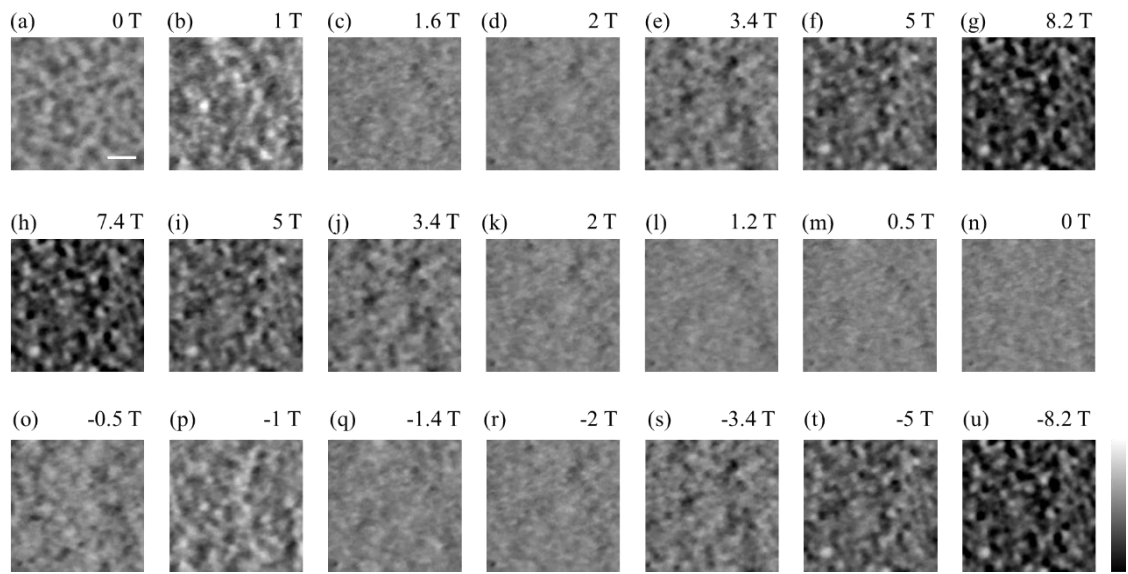

FIG. 2. (a-u) MFM images of sample #3 at selected magnetic fields with field values shown on the upper-right corner. The color scale is from -0.03 Hz to 0.03 Hz. The lateral scale bar is 1 μm.

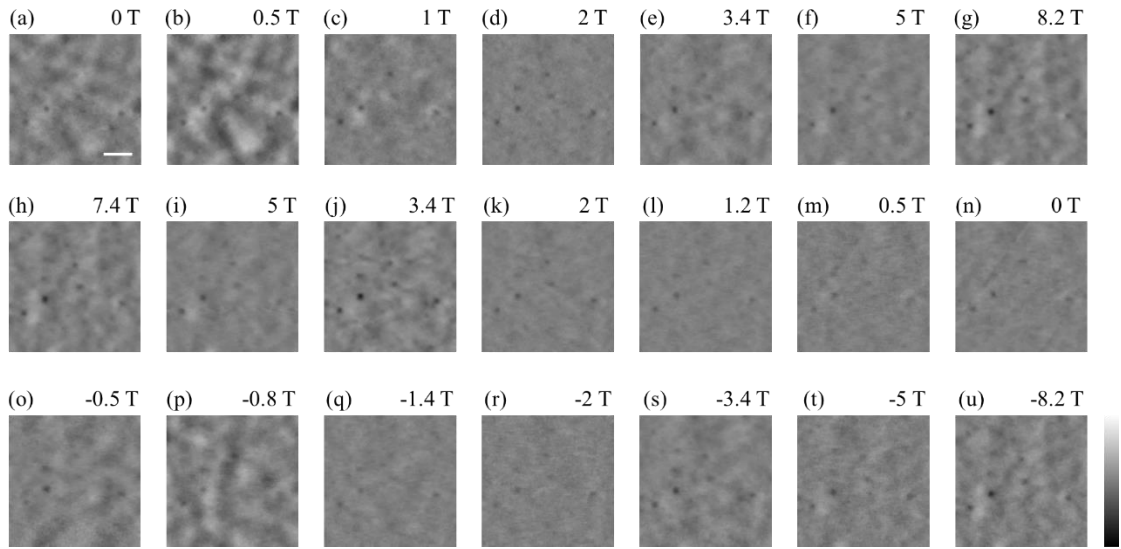

FIG. 3. (a-u) MFM images of sample #5 at selected magnetic fields with field values shown on the upper-right corner. The color scale is from -0.03 Hz to 0.03 Hz. The lateral scale bar is 1 μm.

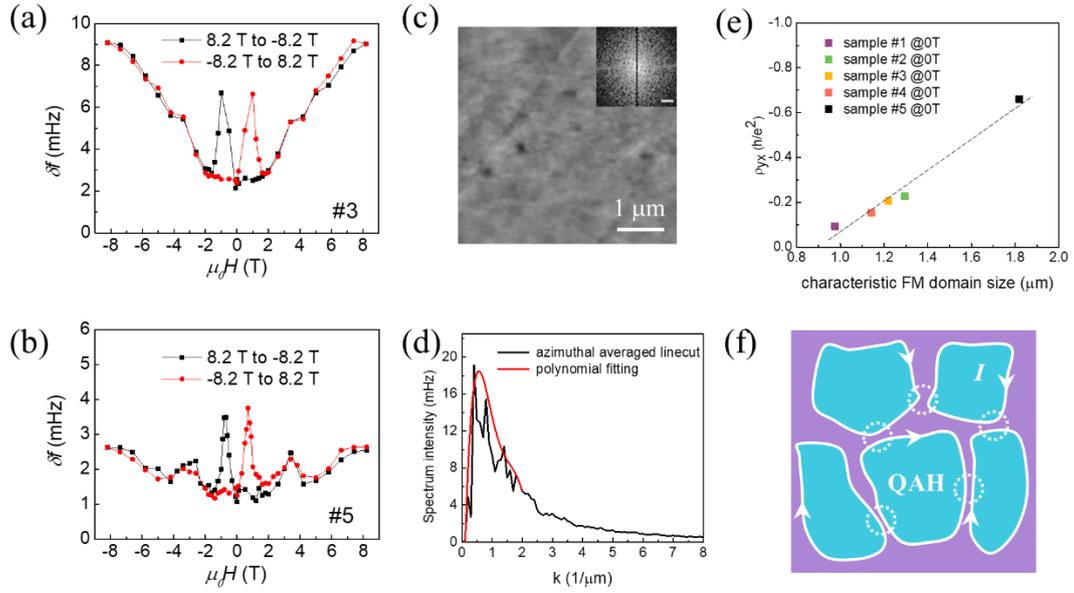

FIG. 4. (a),(b) The field dependence of the standard deviation of the MFM signal from sample #3 and #5, respectively. (c) The zero field MFM image of sample #5 after the initial magnetization and its Fourier transformation (FT) image. The color scale of MFM image is from -0.015 Hz to 0.015 Hz. The scale bar of FT image is 2/μm. (d) $k$ dependence of the angular averaged FT spectrum intensity and the polynomial fitting to its envelope. (e) The zero field Hall resistivity against the estimated magnetic domain size at zero field. The dash line in the figure is guide for the eye. (f) A schematic of the QAH puddles network.